\documentclass{iaus}
\usepackage{graphicx}
\usepackage{graphicx,bm,url}
\usepackage{natbib}

\newcommand{\EQ}{\begin{equation}}
\newcommand{\EN}{\end{equation}}
\newcommand{\EQA}{\begin{eqnarray}}
\newcommand{\ENA}{\end{eqnarray}}

\newcommand{\Eq}[1]{Equation~(\ref{#1})}

\newcommand{\Fig}[1]{Figure~\ref{#1}}

{}
{}
{}

{}
{}
{}
{}
{}
{}
{}
{}
{}
\newcommand{\meanBB}{\overline{\mbox{\boldmath $B$}}{}}{}
{}
{}
{}
{}
{}
{}
{}
{}
{}
{}

{}
{}
{}

{}

{}
{}

\newcommand{\uu}{\mbox{\boldmath $u$} {}}

\newcommand{\oo}{\mbox{\boldmath $\omega$} {}}

\def\ga{\mathrel{\mathchoice {\vcenter{\offinterlineskip\halign{\hfil
$\displaystyle##$\hfil\cr>\cr\sim\cr}}}
{\vcenter{\offinterlineskip\halign{\hfil$\textstyle##$\hfil\cr>\cr\sim\cr}}}
{\vcenter{\offinterlineskip\halign{\hfil$\scriptstyle##$\hfil\cr>\cr\sim\cr}}}
{\vcenter{\offinterlineskip\halign{\hfil$\scriptscriptstyle##$\hfil\cr>\cr\sim\cr}}}}}
\def\Co{\mbox{\rm Co}}
\def\Sh{\mbox{\rm Sh}}

\def\Pm{\mbox{\rm Pr}_M}
\def\Rm{\mbox{\rm Re}_M}

\def\Rmc{R_{\rm m,{\rm crit}}}

\def\Co{\mbox{\rm Co}}

\def\kmean{k_{\rm m}}

\def\kf{k_{\rm f}}
\def\epsf{\epsilon_{\rm f}}

\def\urms{u_{\rm rms}}

\def\etat{\eta_{\rm t}}

\def\etaT{\eta_{\rm T}}

\def\Beq{B_{\rm eq}}

\def\onethird{{\textstyle{1\over3}}}

\newcommand{\yjgr}[3]{ #1, {J.\ Geophys.\ Res.,} {#2}, #3}

\newcommand{\yapj}[3]{ #1, {ApJ,} {#2}, #3}

\newcommand{\yapjl}[3]{ #1, {ApJ,} {#2}, #3}
\newcommand{\yapjs}[3]{ #1, {ApJS,} {#2}, #3}

\newcommand{\yan}[3]{ #1, {Astron.\ Nachr.,} {#2}, #3}

\newcommand{\yana}[3]{ #1, {A\&A,} {#2}, #3}

\newcommand{\ygafd}[3]{ #1, {Geophys.\ Astrophys.\ Fluid Dyn.,} {#2}, #3}

\newcommand{\yjfm}[3]{ #1, {J.\ Fluid Mech.,} {#2}, #3}

\newcommand{\ypf}[3]{ #1, {Phys.\ Fluids,} {#2}, #3}

\newcommand{\ypp}[3]{ #1, {Phys.\ Plasmas,} {#2}, #3}

\newcommand{\yjetp}[3]{ #1, {Sov.\ Phys.\ JETP,} {#2}, #3}

\newcommand{\yprl}[3]{ #1, {Phys.\ Rev.\ Lett.,} {#2}, #3}

\newcommand{\ymn}[3]{ #1, {MNRAS,} {#2}, #3}
\newcommand{\ynat}[3]{ #1, {Nature,} {#2}, #3}

\newcommand{\ysph}[3]{ #1, {Solar Phys.,} {#2}, #3}

\newcommand{\ypre}[3]{ #1, {Phys.\ Rev.\ E,} {#2}, #3}

\newcommand{\yjour}[4]{ #1, {#2}, {#3}, #4}

\newcommand{\ybook}[3]{ #1, {#2} (#3)}

\title[From convective to stellar dynamos]
{From convective to stellar dynamos}

\author[A.\ Brandenburg, P.\ J.\ K\"apyl\"a, \& M.\ J.\ Korpi]
{Axel Brandenburg$^{1,2}$, Petri J.\ K\"apyl\"a$^{1,3}$, Maarit J.\ Korpi$^{3}$
}

\affiliation{
$^1$NORDITA, Roslagstullsbacken 23, SE-10691 Stockholm, Sweden\\
$^2$Department of Astronomy, Stockholm University, SE-10691 Stockholm, Sweden\\
$^3$Department of Physics, PO Box 64, FI-00014 University of Helsinki, Finland
}

\pubyear{2011}
\volume{271}
\pagerange{1--9}
\date{$ $Revision: 1.25 $ $}
\jname{Astrophysical Dynamics: From Stars to Galaxies}
\editors{N. Brummell, A.S. Brun, M.S. Miesch, \& Y. Ponty, eds.}
\begin{document}

\maketitle

\begin{abstract}
Convectively driven dynamos with rotation
generating magnetic fields on scales large compared
with the scale of the turbulent eddies are being reviewed.
It is argued that such fields can be understood as the result of an
$\alpha$ effect.
Simulations in Cartesian domains show that such large-scale magnetic fields
saturate on a time scale compatible with the resistive one, suggesting
that the magnitude of the $\alpha$ effect is here still constrained by
approximate magnetic helicity conservation.
It is argued that, in the absence of shear and/or any other known
large-scale dynamo effects, these simulations prove the existence of
turbulent $\alpha^2$-type dynamos.
Finally, recent results are discussed in the context of solar and stellar
dynamos.
\keywords{MHD -- turbulence -- Sun: magnetic fields}
\end{abstract}

\firstsection

\section{Introduction}

Stars with outer convection zones are known to display magnetic activity,
often in a cyclic fashion like in the Sun.
Such activity can generally be explained by a turbulent dynamo influenced
by rotation and stratification to produce the anisotropies required
for the generation of large-scale magnetic fields.
The basic theory is reasonably well understood \citep{Mof78,Par79,KR80},
but there continues to be substantial controversy until the present day.
A major stumbling block has been the understanding of what is known
as catastrophic quenching \citep{VC92,CH96} and resistively limited
saturation \citep{B01}, as well as the very existence of the
$\alpha$ effect in convection even without nonlinearity \citep{CH06,HC08}.

The first two issues have been reviewed in detail by \cite{BS05}.
The purpose of the present paper is to review recent progress concerning
convective dynamos.
However, in view of applications to solar and stellar dynamos,
it is important to realize that we are still lacking simulations
that reproduce the salient features of the solar dynamo.
We should therefore keep our eyes open for new phenomena that may emerge
as simulations become more realistic.

\section{Excitation conditions of small-scale and large-scale dynamos}

\begin{figure}[t!]\begin{center}
\includegraphics[width=.49\textwidth]{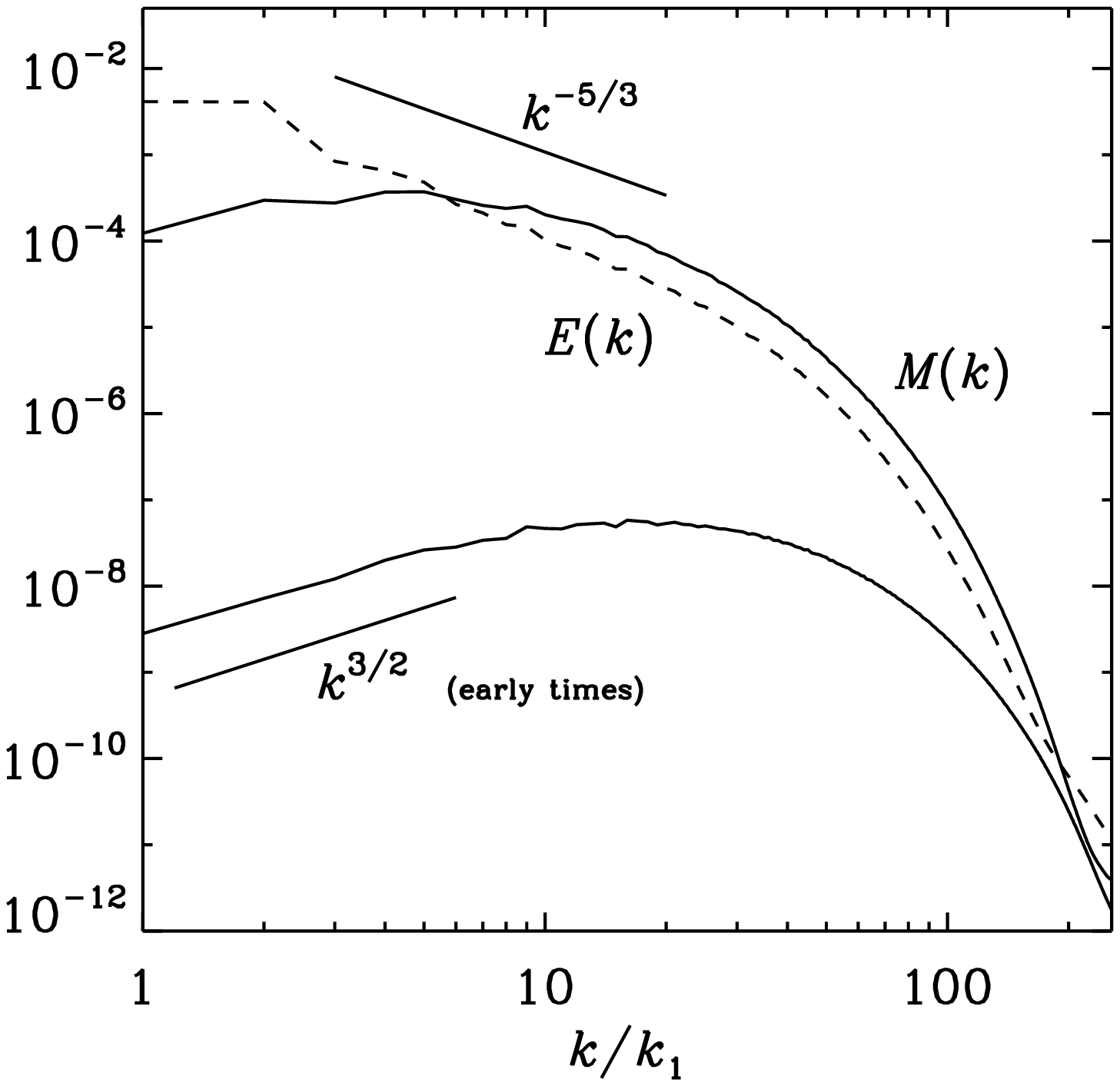}
\includegraphics[width=.49\textwidth]{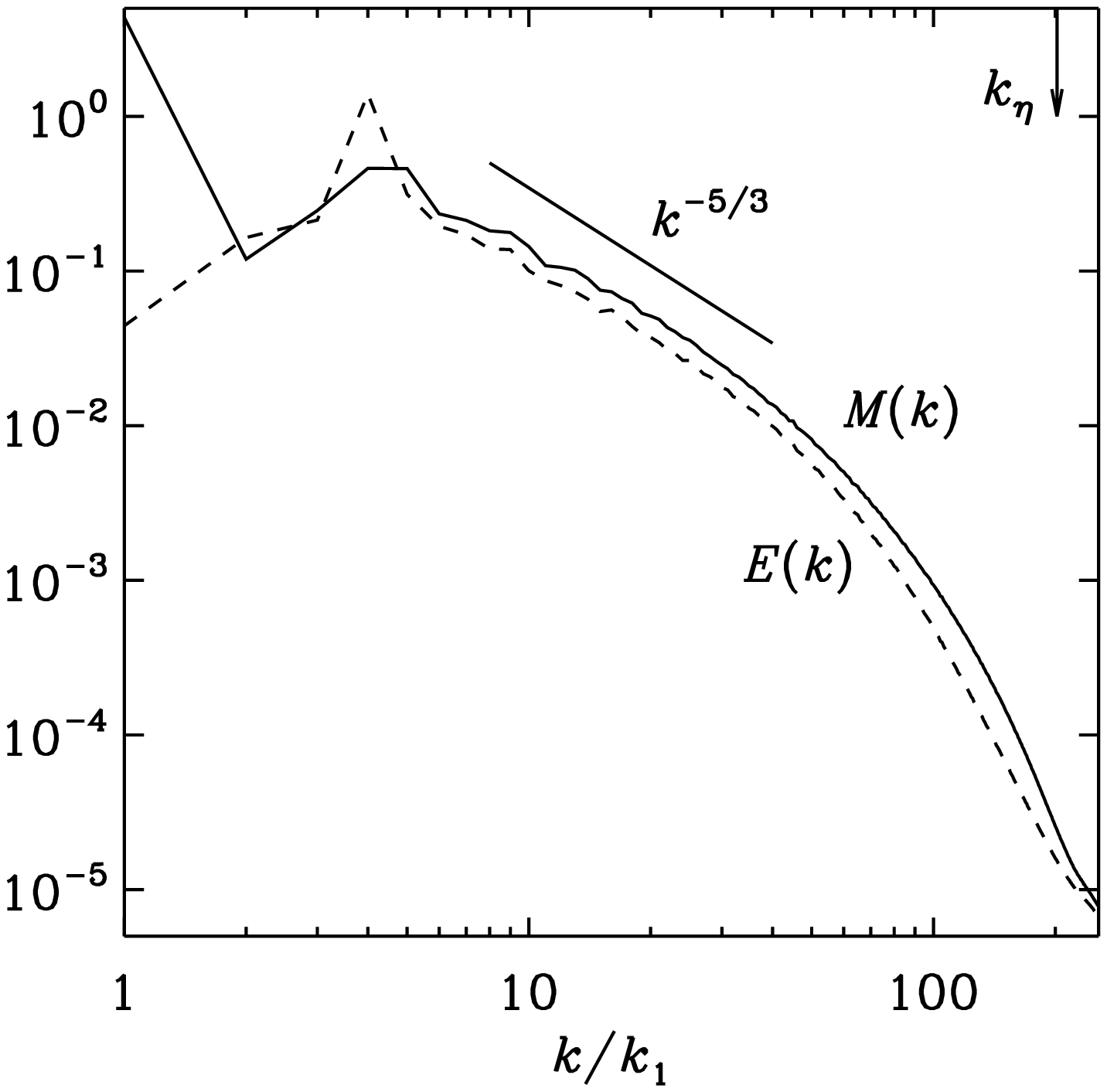}
\end{center}\caption[]{
Kinetic and magnetic energy spectra in a turbulence simulation without
net helicity (left) and with net helicity (right) for a magnetic Prandtl
number of unity and a mesh size is $512^3$ meshpoints.
Notice the pronounced peak of $M(k)$ at $k=k_1$ in the case with helicity.
Adapted from \cite{BS05} and \cite{B09}, respectively.
}\label{pspec_nohel512d2_pspec_PrM1}\end{figure}

Small-scale and large-scale dynamos are quite different in nature.
The difference becomes most evident in the nonlinearly saturated regime,
provided one allows for what we call scale separation, which means
that the size of the domain is large compared with the scale of the
largest (energy-carrying) eddies of the turbulence.
In \Fig{pspec_nohel512d2_pspec_PrM1} we show spectra highlighting
the remarkable difference between the two types of dynamos.
Conversely, if there is insufficient scale separation, a large-scale
dynamo becomes impossible and both types of simulations would look very
similar, as has been demonstrated by \cite{HBD04}; see their Fig.~23.

There can be different types of large-scale dynamos.
The most frequently studied ones are the $\alpha^2$ and $\alpha\Omega$
type dynamos.
These are dynamos that can produce large-scale magnetic fields owing
to the presence of kinetic helicity in the turbulence, giving rise to
an $\alpha$ effect.
The presence of shear can further modify the dynamo, making it usually easier
to excite
and favoring oscillatory over non-oscillatory solutions.

Shear can be a typical result of rotation of a gaseous body in
the presence of anisotropic turbulence \citep{Rue80,Rue89}.
Shear alone is often found to give rise to large-scale fields -- even
if the turbulence is non-helical \citep{B05,Y08a,Y08b,BRRK08}.
The nature of such dynamo action is still a matter of debate and ranges
from {\em incoherent} $\alpha\Omega$ dynamos \citep{VB97,Pro07} to
shear--current dynamos \citep{RK03,RK04}.

Returning to the $\alpha^2$ and $\alpha\Omega$ dynamos, it is important
to realize that their excitation conditions are generally quite different.
The onset of small-scale dynamo action depends generally on the value of
the magnetic Reynolds number,
\EQ
\Rm=\urms/\eta\kf,
\EN
where $\urms$ is the typical rms velocity of the turbulence, $\eta$
is the microscopic magnetic diffusivity, and $\kf$ is the forcing
or integral wavenumber, i.e.\ the wavenumber of the energy-carrying
motions.
This is roughly where the peak of the energy spectrum is located.
The critical value, $\Rmc$, above which dynamo action commences, depends
on the value of the magnetic Prandtl number, $\Pm=\nu/\eta$, where
$\nu$ is the microscopic kinematic viscosity and is about $35$ for
$\Pm=1$ \citep{NRS83,Sub99,HBD04}, but increases to values around and
above 400 for $\Pm$ somewhere between $0.2$ and $0.1$ \citep{Scheko05}.
There is now also evidence that $\Rmc$ may actually show a peak at
$\Pm=0.1$ and might then drop to slightly lower values for $\Pm=0.05$
and below \citep{Iska07}.
This unusual behavior is connected with a change of the ``roughness''
of the velocity field \citep{BC04} and the occurrence of a bottleneck
effect in the velocity spectrum \citep{Fal94,DHYB03}, which means that
the velocity has maximum roughness for $\Pm\approx0.1$ when the
resistive scale coincides with the position of the bottleneck.

The situation is quite different with large-scale dynamos that operate
completely independently of the value of $\Pm$, as long as $\Rm$
is large enough.
Already in \cite{B01} the critical value of $\Rm$ was found to be around
unity regardless of whether $\Pm=1$ or 0.1.
This finding was then extended by \cite{Min07} and \cite{B09}, who
demonstrated dynamo action down to $\Pm=0.005$ and 0.001, respectively,
or up to $\Pm=1000$ \citep{B11}.
The conclusion is that large-scale dynamo action depends solely on the
{\em dynamo number}, which is given by $D=C_\alpha$ for $\alpha^2$ dynamos
and by $D=C_\alpha C_S$ for $\alpha\Omega$ (or $\alpha$--shear) dynamos.
Here,
\EQ
C_\alpha=\alpha_0/\etaT k_1\quad\mbox{and}\quad C_S=S/\etaT k_1^2,
\EN
where $\etaT=\etat+\eta$ is the sum of turbulent and microscopic magnetic
diffusivities, $\alpha_0$ is a typical value of the $\alpha$ effect, and
$S$ is the shear rate (i.e.\ a typical value of the velocity gradient).
For $C_\alpha$ and $C_S$, we use standard estimates:
\EQ
\alpha_0\approx-{\tau\over3}\overline{\oo\cdot\uu}
\approx-{\epsf\over3\urms\kf}\kf\urms^2
=-\onethird\epsf\urms,
\EN
where $\epsf=\overline{\oo\cdot\uu}/\kf\urms^2$  is a measure of the
relative kinetic helicity, $\tau=(\urms\kf)^{-1}$ is the turnover time,
and
\EQ
\etat\approx{\tau\over3}\overline{\uu^2}\approx\urms/3\kf.
\EN
With this we find
\EQ
C_\alpha=-{\onethird\epsf\urms/k_1\over\urms/3\kf+\eta}
=-\iota\epsf{\kf\over k_1}
\EN
where
\EQ
\iota=1/\left(1+3\Rm^{-1}\right)
\EN
is a correction factor that is close to unity for $\Rm\gg1$.
Furthermore, we have
\EQ
C_S={S/k_1^2\over\urms/3\kf+\eta}
={3\iota S\over\urms\kf}\left({\kf\over k_1}\right)^2
=3\iota\Sh\left({\kf\over k_1}\right)^2,
\EN
where we have defined the shear parameter $\Sh=S/\urms\kf$.
Note that, especially in the presence of shear, the possibility of dynamo
action is strongly connected with the scale separation ratio.
Indeed,
\EQ
D=-3\iota\epsf\Sh\left({\kf\over k_1}\right)^3
\EN
depends cubicly on the scale separation ratio.
This explains why $\alpha\Omega$ dynamos are often much easier to obtain
than $\alpha^2$ dynamos.

\begin{figure*}[t]
\centering
\includegraphics[width=\textwidth]{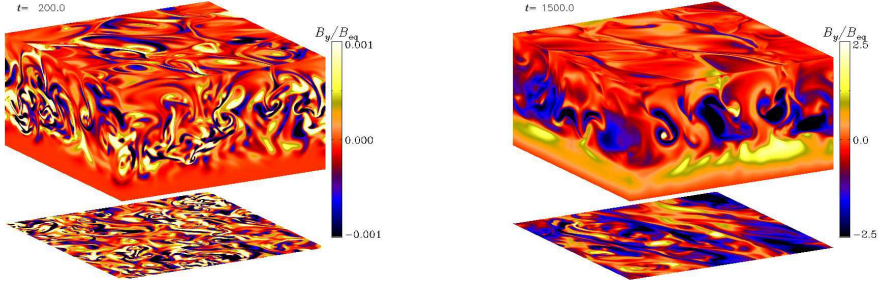}
\caption{Snapshots of $B_y$ in the early phase ({\it left}) and
saturated phase ({\it right}) of a convective dynamo with shear.
Adapted from \cite{KKB08}; see also 
\texttt{http://www.helsinki.fi/\ensuremath{\sim}kapyla/movies.html}}
\label{100+2000}
\end{figure*}

\section{Large-scale dynamos in Cartesian domains}

\subsection{Dynamos of $\alpha\Omega$ type}

Given the alarming signs of earlier investigations by \cite{CH06} and
\cite{HC08}, it was quite unclear whether the $\alpha$ effect even
exists in simulations with convection.
At the time, several possible reasons were put forward, including
the absence of stratification; see, for example, \cite{B09_bern}.
In the wake of this initial frustration, it was quite surprising
when large-scale dynamo action was found in rotating convection in
the presence of shear \citep{KKB08}; see \Fig{100+2000}.
Similar results were obtained by \cite{HP09}.
This controversy was still ongoing at the conference
``Turbulence and Dynamos'' in Stockholm in March 2008\footnote{\url
{http://agenda.albanova.se/conferenceDisplay.py?confId=325}
} where Hughes\footnote{\url
{http://videos.nordita.org/conference/Turbulence2008/hires/March17/Part1.WMV}
} argued that no convective large-scale dynamos exist,
while K\"apyl\"a\footnote{\url
{http://videos.nordita.org/conference/Turbulence2008/hires/March19/Part5.WMV}
} showed results from low Reynolds number convection with shear 
where large-scale fields were indeed obtained.
In an attempt to clarify the still conflicting results regarding the actual
value of $\alpha$ in rotating convection, \cite{KKB10a} pointed out that for
a nonuniform mean field, the mean current density cannot be neglected.
In this case, the turbulent magnetic diffusivity contributes and
explains the small net electromotive force measured by \cite{HP09} by
imposing a uniform field.

\subsection{Dynamos of $\alpha^2$ type}

The simulations mentioned above do not provide conclusive evidence
for the existence of an $\alpha$ effect in rotating convection,
because it is in principle possible that the dynamo could be the
result of an incoherent $\alpha\Omega$ dynamo or a shear--current dynamo.
In the absence of shear, however, there is no viable alternative to
an $\alpha^2$ dynamo.
It is therefore important to consider the conceptually simpler case
without imposed shear, as was also emphasized by \cite{HPC11}, who
noted that this was not done by \cite{KKB10a}, who just considered
the case of a sinusoidal shear profile.
For this reason, we discuss in the following the papers of \cite{KKB09a}
and, in particular, \cite{KKB09b}, where large-scale dynamo action was
studied in non-shearing convection at sufficiently large Coriolis numbers.

Before trying to simulate an $\alpha^2$ dynamo for rotating convection,
it is instructive to obtain guidance from numerically obtained measurements
of the $\alpha$ and turbulent diffusivity tensors.
This can be done using the test-field method \citep{Sch05,Sch07}, which
has been applied to turbulence in a number of recent papers
\citep{B05_QPO,BRRK08}.
The result is shown in \Fig{palpeta}.
Using this method, \cite{KKB09a} noted that the magnitude of the
relevant components of
the $\alpha$ tensor vary only weakly with Coriolis number,
$\Co=2\Omega/\urms\kf$, where $\Omega$ is angular velocity,
while $\etat$ diminishes with increasing values of $\Co$.
This was a clear indication that dynamos of $\alpha^2$ type might become
possible once $\Co$ is large enough.
We emphasize this point, because it is one of the several examples where
mean-field theory has proven its predictive power.

\begin{figure}[t]
\begin{minipage}[b]{0.5\linewidth}
\centering
\includegraphics[width=\columnwidth]{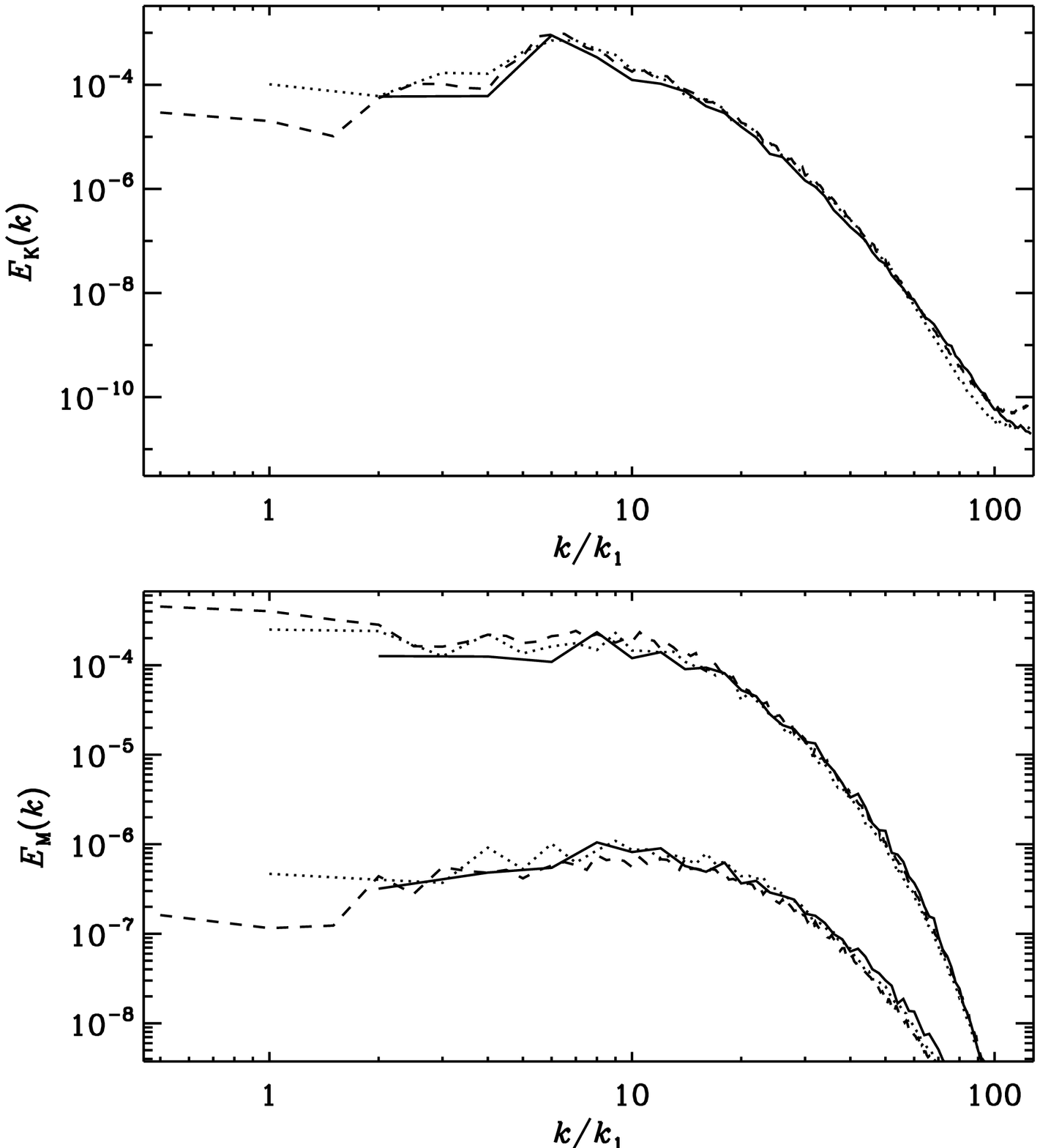}
\caption{Two-dimensional power spectra of velocity (upper panel) and
  magnetic field (lower panel) as functions of system size.
  In the lower panel
  the upper curves show the spectra from the saturated state whereas the
  lower curves show the spectra from the kinematic state multiplied by $10^7$.
  Adapted from \cite{KKB09b}.
}\label{pspec_size}
\end{minipage}
\hspace{0.5cm}
\begin{minipage}[b]{0.5\linewidth}
\centering
\includegraphics[width=\columnwidth]{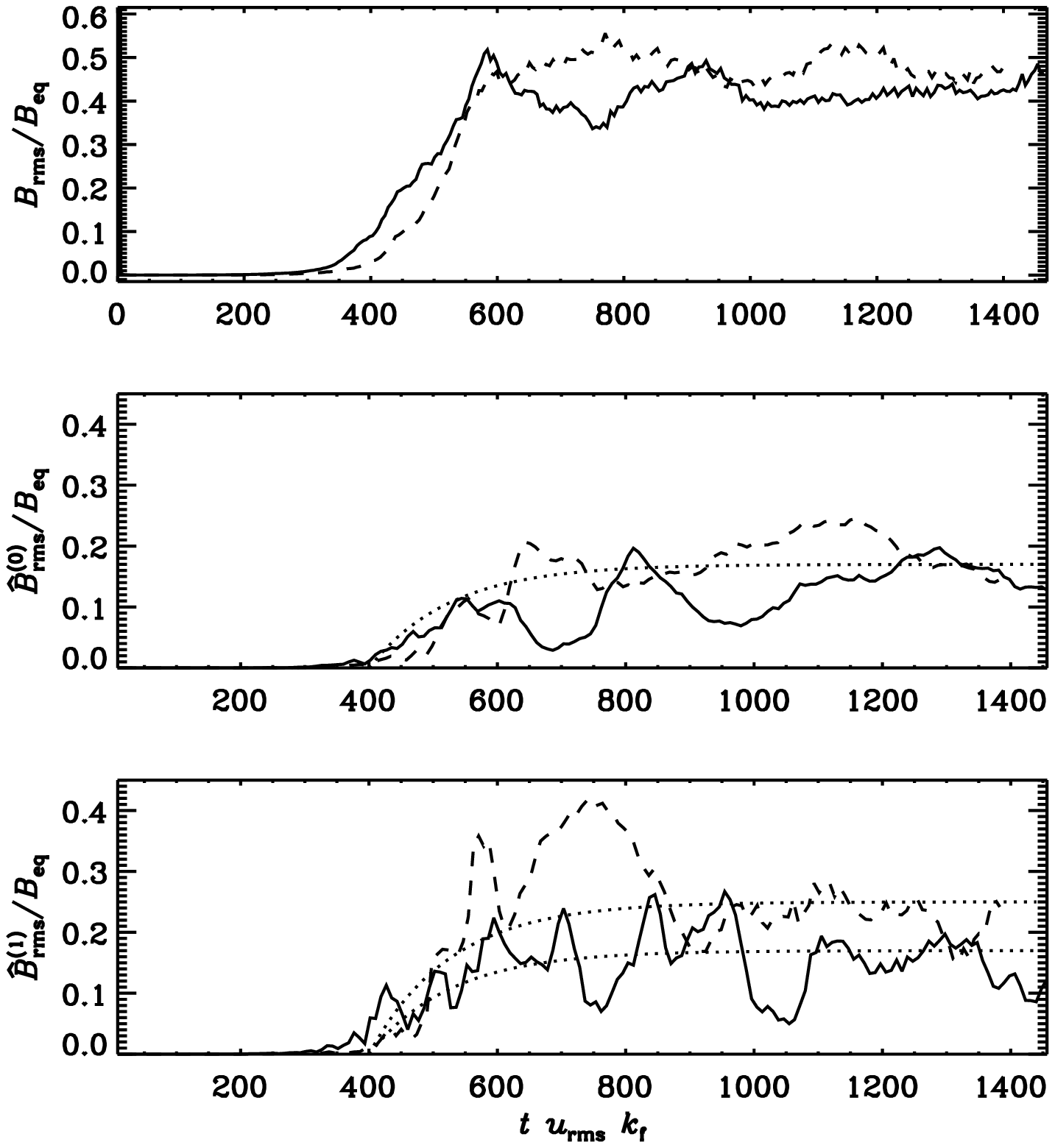}
\caption{Top panel: rms-values of the total magnetic field as
  functions of time for vertical field (solid lines)
  and perfect conductor boundary conditions (dashed
  lines). The two lower panels show the sums of the rms-values of
  the Fourier amplitudes of $B_x$ and $B_y$ for $k/k_1=0$ (middle
  panel) and $k/k_1=1$ (bottom panel). The dotted lines in the two
  lower panels show a saturation predictor according to the model of
  Brandenburg (2001).
  Adapted from \cite{KKB09b}.
}\label{psatb}
\end{minipage}
\end{figure}

\begin{figure}[t]
\centering
\includegraphics[width=\columnwidth]{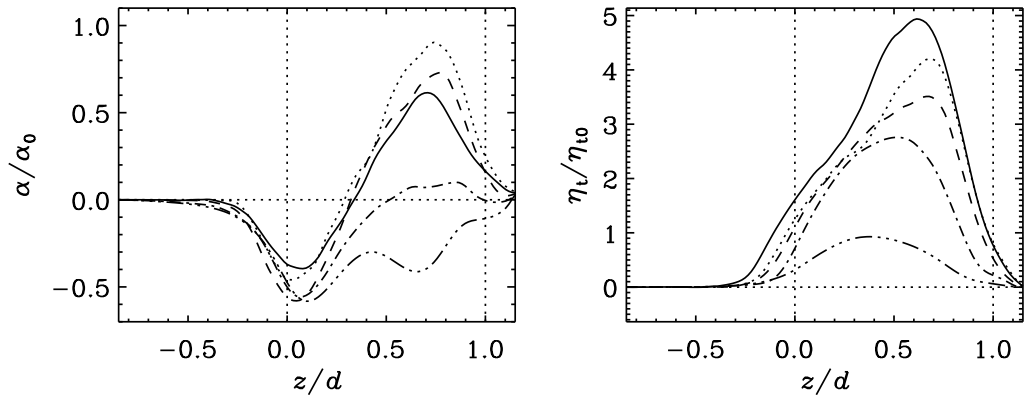}
\caption{Normalized profiles of $\alpha$ (left panel) and $\etat$ 
  (right panel) from
  kinematic test field simulations. The vertical dotted lines at
  $z/d=0$ and $z/d=1$ indicate the base and top of the convectively
  unstable layer, respectively.
  Adapted from \cite{KKB09b}.
}\label{palpeta}
\end{figure}

Consequently, in a subsequent investigation, \cite{KKB09b} carried out
simulations for large enough values of $\Co$ and did indeed find
dynamo action of large-scale type when $\Co\ga10$.
The large-scale field became even more pronounced as the aspect ratio
was increased.
In \Fig{pspec_size} we present horizontal spectra of magnetic and
kinetic energies.
What is important here is the fact that, even though the magnetic energy
is less (by factor 5) than the kinetic energy at what we estimate to be
$\kf$ (about $5k_1$), the magnetic energy strongly {\em exceeds} the
kinetic energy at the scale of the domain.
This seems to exclude alternative explanations whereby the magnetic field
at smaller wavenumbers might just be a trivial result of diffusion in
wavenumber space.
Instead, we argue that this is strong evidence for the physical reality
of an $\alpha^2$ dynamo driven by rotating convection.

In agreement with virtually all earlier work on large-scale dynamos of
$\alpha^2$ type the saturation time of the dynamo is comparable with the
resistive time.
Indeed, \cite{B01} found that in the absence of strong magnetic helicity
fluxes, the saturation of an $\alpha^2$ dynamo follows a switch-on behavior
where, after saturation, the mean field is given by
\EQ
{\meanBB^2\over\Beq^2}\approx
{\kf\over\kmean}\left[1-e^{-2\eta\kmean^2(t-t_{\rm s})}\right].
\label{SatFit}
\EN
This is also seen in the present case; see \Fig{psatb}, where we overplot
the prediction from \Eq{SatFit}.

It is likely that diffusive magnetic helicity fluxes are present in the
convection simulations discussed above \citep{BCC09}.
Those fluxes could in principle give rise to faster saturation times
than what is seen in \Fig{SatFit}.
This question has been addressed quantitatively by \cite{Mitra10} and
\cite{HB10}, who note that at the magnetic Reynolds numbers accessible
so far, diffusive helicity fluxes are still quite weak compared with
the resistive processes.
Based on their scalings for different values of $\Rm$, they estimate
that resistive saturation effects would only begin to be alleviated
for $\Rm$ well in excess of values around 1000 or even $10^4$;
see also Fig.~10 of \cite{CHBM11}.
Convection simulations with closed magnetic boundaries do seem to
suffer from catastrophic quenching at least up to $\Rm\approx240$
\citep{KKB10b}.
Reaching much higher values to study this issue
further in the near future is not possible.
While this is certainly somewhat discouraging news, it does highlight
the importance of studying detailed scaling properties of large-scale
dynamos rather than producing solitary examples of dynamos at large resolution,
hoping that they represent the Sun.

\section{Bi-helical magnetic field}

An important property of $\alpha$ effect dynamos is the fact that they produce
bi-helical magnetic fields.
This means that one expects to see magnetic helicity fluxes with opposite
signs at small and large scales.
While this is now well established theoretically \citep{BB03,YB03},
there is still no widespread observational evidence for this.
Large-scale magnetic helicity can be estimated using synoptic maps of
the azimuthally averaged radial magnetic field of the Sun; see Fig.~1
of \cite{BBS03}.
Small-scale magnetic helicity fluxes have been inferred from
magnetic field measurements in active regions
and coronal mass ejections \citep{BR00}.

There is also some evidence from measurements of magnetic helicity in
the solar wind.
Using the assumption of homogeneity, \cite{MGS82} were able to determine
magnetic helicity spectra for the solar wind.
Preliminary analysis of more recent solar wind data from the Ulysses
spacecraft does indeed suggest that the field in the solar wind is
also bi-helical.
Further details on this are presented in a dedicated publication
\citep{BSBG11}.

\section{Concluding remarks}

We still do not know exactly how the solar dynamo works.
If it is of $\alpha\Omega$ type, given that the $\alpha$ effect
is positive in the northern hemisphere, and using the fact that
radial shear is positive in the bulk of the convection zone,
one would expect poleward migration of the dynamo wave.
This is also what three-dimensional simulations in spherical
shells have shown repeatedly over several decades starting with
the early work of \cite{Gil83}, and now again in the spherical
wedge simulations of \cite{Kapy10}.
For rapid rotation, however, polarity flips of toroidal magnetic
field can also occur more abruptly, as has been demonstrated by
\cite{Brown10} and \cite{GCG10}, which is beginning to be reminiscent
of polarity reversals in the geodynamo \citep{GR95}, but is different
from what we know about the solar dynamo.
In this connection it is worth recalling yet another
recent surprise: oscillatory solutions with equatorward migration are
also possible in the absence of any differential rotation provided
the dynamo is somehow bounded between highly conducting media at high
latitudes \citep{Mitra10b}.
It is obviously unclear whether this has any bearing on the solar
dynamo, but it reminds us of the possibility of surprises.

Other proposals for how the solar dynamo might work include the flux
transport dynamo \citep{Dur95,CSD95,DG99}, and the possibility of a
dynamo shaped by the negative radial angular velocity gradient in
the near-surface shear layer \citep{B05}.
Neither of these two scenarios have been seen in three-dimensional turbulence
simulations.
The former suffers from the difficulty of obtaining a sufficiently
coherent meridional circulation that does not break up into smaller
circulation patterns, while the latter may suffer from the difficulty
of resolving the small-scale turbulence in the near-surface shear layer.
A possible step forward might therefore be a combined effort utilizing
a range of different simulations in spherical and Cartesian geometries
on the one hand, and improved mean-field theory on the other.
Clearly, in order to improve mean-field theory it is essential to seek
guidance from direct simulations, as has already been done in
recent years with increasing success.

\acknowledgments
We acknowledge the allocation of computing resources provided by the
Swedish National Allocations Committee at the Center for
Parallel Computers at the Royal Institute of Technology in
Stockholm and the National Supercomputer Centers in Link\"oping
as well as the Norwegian National Allocations Committee at the
Bergen Center for Computational Science and the computing facilities
hosted by CSC - IT Center for Science Ltd.\ in Espoo, Finland,
who are administered by the Finnish Ministry of Education.
This work was supported in part by
the European Research Council under the AstroDyn Research Project
227952, the Swedish Research Council grant 621-2007-4064, and the
Finnish Academy grants 121431, 136189, and 112020.


\begin{discussion}

\discuss{M. Proctor}{ 
What do you do about the gauge when calculating magnetic helicity, and
does it make any difference to the answers obtained?
}

\discuss{A. Brandenburg}{
Yes, the magnetic helicity density is in general gauge-dependent.
However, if there is sufficient scale separation between mean and fluctuating
fields, the magnetic helicity density computed from the fluctuating field
can be shown to be gauge-invariant \citep{SB06}.
\cite{HB10} have recently confirmed this in a simulation where the magnetic
helicity from the mean field was strongly gauge dependent, but that from
the fluctuating field was not.
}

\discuss{T. Rogers}{ 
To calculate magnetic helicity you assumed that the solar wind was isotropic,
which observations show it is not. How would this affect the results
you present?
}

\discuss{A. Brandenburg}{
Since we have to adopt the Taylor hypothesis, only the two
magnetic field components perpendicular to the radial direction
enter the calculation.
The field in the plane perpendicular to the radial direction is still
fairly isotropic, so I guess our results are still meaningful.
To clarify the significance of our results further, it might be useful
to compute magnetic helicity from simulations of anisotropic MHD turbulence
with one preferred direction.
}

\discuss{D. Hughes}{ 
In your simulations that show the generation of large-scale fields on
a long time, what is the timescale for the generation of the fields?
If it is ohmic then it is not surprising.
}

\discuss{A. Brandenburg}{ 
The initial exponential growth occurs always on a dynamical time scale,
but full saturation is only obtained on a resistive time scale.
We know that this problem can only be alleviated by magnetic helicity
fluxes, which are quite weak in our Cartesian simulations.
Nevertheless, our simulations prove the point that the $\alpha$ effect
works in rotating convection, which was until now quite unclear.
}

\discuss{C. Forest}{ 
How do you model the boundary conditions? Open or Closed?
}

\discuss{A. Brandenburg}{ 
At the bottom we adopt a perfect conductor boundary condition and at the
top we assume that the horizontal field vanishes.
This pseudo-vacuum condition is numerically more robust than a proper
vacuum condition.
However, it would be more realistic to couple the convection simulation
to a force-free model, as has been done for forced turbulence simulations
in the paper by \cite{WB10}, which is also presented here as a poster.
}

\end{discussion}
\end{document}